# ETUDE D'UN MODELE D'INVERSION LIANT L'EMISSIVITE A L'HUMIDITE DES SOLS. CONTRIBUTION A LA MISE AU POINT DE L'ALGORITHME DE LA MISSION SMOS

F. Demontoux[1], B. Le Crom[1,2], G. Ruffié[1], JP. Wigneron[2], J.P. Grant[2,3] Heather Lawrence[1]

[1] Laboratoire PIOM-ENSCPB-UMR 5501- 16 av Pey-Berland 33607 Pessac
[2] INRA-Unité de Bioclimatologie, BP 81, Villenave d'Ornon Cedex 33883
[3] Faculty of Earth and Life Sciences, Vrije Universiteit Amsterdam, De Boelelaam 1085.

## 1   Introduction

SMOS (Soil Moisture and Ocean Salinity), dont le lancement est prévu pour l'horizon 2007, est la seconde mission d'opportunité « Earth Explorer » à être développée dans le cadre du programme « Living Planet » de l'agence spatiale européenne (ESA)[1]. Sur la problématique du cycle de l'eau, les données acquises par SMOS permettront d'établir une carte spatiale de l'humidité des surfaces continentales et de la salinité de la couche superficielle des océans. Les applications sont multiples. Sur Terre, la rétention d'eau dans le sol joue un rôle primordial dans l'évolution climatique car l'humidité des sols représente une variable clé régulant l'échange d'eau et d'énergie thermique entre la terre et l'atmosphère. En mer, la salinité est un paramètre qui influence la circulation des masses d'eau dans les océans et entraîne la formation de phénomènes climatiques tel qu'El Niño.

Installé sur la plateforme Protéus, ce satellite, contenant le tout premier radiomètre interférométrique 2D, effectuera donc la 1$^e$ cartographie à l'échelle planétaire de l'humidité des sols et de la salinité des océans et ce grâce à un unique instrument de mesure capable de capture d'images des radiations micro ondes émise autour de 1.4GHz. Les hyperfréquences sont sensibles aux changements de la constante diélectrique du milieu et donc toute variation de la quantité d'eau induit des modifications des propriétés du diélectrique. Cela affecte l'émissivité, et par conséquent la température de brillance détectée par le radiomètre. Ainsi il existe une relation directe entre l'humidité du sol pour des profondeurs de 2 à 5 cm, la salinité des océans et les émissions d'origine terrestre sur la fréquence de 1.4 GHz.

En raison de la résolution spatiale du pixel de SMOS, de l'ordre de 40 Km sur 40 Km, la tache au sol de la mesure englobe généralement un grand nombre de type d'occupation du sol. Les forêts sont présentes dans une majorité des pixels en zone tropicale, boréale et tempérée. Les forêts sont des couverts relativement opaques, sur lesquels le suivi de l'humidité reste problématique. En particulier, l'effet de la litière a, jusqu'ici, été négligé.

Le but de notre travail a été dans un premier temps de réussir à mesurer les propriétés diélectriques d'un type de litière et de terre afin d'intégrer ces valeurs à un modèle analytique multi couches de sol [4]. L'objectif est de mettre en évidence les effets de cette strate sur le système multi couche global. Ceci permettra d'aboutir à une formulation analytique simple d'un modèle de litière qui puisse être intégré à l'algorithme de calcul de SMOS afin de recueillir des informations sur l'humidité à partir des mesures d'émissivité. La méthode de mesure que nous avons utilisé [4] permet de présenter les résultats sous forme de domaines de permittivité qui intègreront les erreurs de mesures et de répétitivité des mesures.

Jusqu'à présent la présence de la litière a été négligée. L'effet de la végétation (arbre et sous bois) est alors caractérisé par son épaisseur optique τ et son albédo de simple diffusion. Son effet peut être corrigé par l'intermédiaire d'un modèle radiatif simple. Néanmoins les premières simulations laissent penser que l'effet de la litière sur l'émissivité d'un système litière + sol est loin d'être négligeable. Il est donc nécessaire de mettre en place un modèle analytique qui permettrait de corriger l'effet de cette (ces) couche(s) supplémentaire(s).

Le but de cet article est de présenter les modèles analytiques que nous avons retenus pour corriger l'effet de la végétation et de la litière afin de connaître l'émissivité du sol nu. Nous avons développé un modèle numérique (avec le logiciel HFSS) de calcul de l'émissivité de systèmes multicouches [réf] afin de valider les résultats des modèles d'inversion. Des domaines de permittivité seront introduits au modèle afin de tenir compte des perturbations liées à la mesure et de la variation de la teneur en eau des couches.

## 2   Présentation des modèles

Le milieu d'étude que nous considèrerons dans cet article est la forêt. Ce système peut être représebté par un système tri-couche constitué de végétation (les arbres, les sous bois), de litière (débris végétaux) et de terre (sol).

### 2.1   Correction de l'effet de la végétation

L'algorithme développé dans le cadre de la mission SMOS permet de relier la grandeur géophysique recherchée SM (pour « Soil Moisture ») à la grandeur radiométrique mesurée $T_B$ (pour Température de Brillance) et ce par l'intermédiaire d'un modèle simplifié d'équation de transfert radiatif appelé le modèle τ-ω. Ce modèle considère le milieu comme un système bi-couche constitué de la végétation (arbre et sous bois) et du sol (terre + litière). La théorie du transfert radiatif traduit les modifications d'une onde électromagnétique, en intensité et en phase, à la traversée d'un milieu (ici une couche de végétation) en tenant compte de trois phénomènes présents dans le milieu. Le premier phénomène est l'absorption que nous décrirons à l'aide du coefficient $k_a$. Le phénomène de diffusion quand à lui sera intégré par le coefficient $k_s$. Enfin le phénomène d'extinction sera pris en compte à travers le coefficient $k_e=k_a+k_s$.

Grâce à l'approche simplifiée de l'équation de transfert radiatif qui représente le bilan d'énergie radiative [2] il est possible d'exprimer simplement l'émission d'un couvert. Le radiomètre mesure une température de brillance $T_B$ qu'il est possible de relier à l'émissivité globale du couvert par l'intermédiaire de l'approximation de Rayleigh-Jeans. $T_B$ est la somme des énergies sortantes, elle se met alors sous la forme suivante :

$$T_B = T_{B1} + T_{B2} + T_{B3}$$
$$= (1-\omega)(1-\gamma)(1+\gamma\Gamma_S)T_V + \gamma e T_S$$

où   $e$ est l'émissivité du sol ;

$T_S$ est la température effective du sol.

$\Gamma_S$ est la réflectivité du sol et $\Gamma_v$ celle de la végétation que nous négligerons dans un premier temps

$\omega$ est l'albédo de simple diffusion de la végétation ;

$\gamma = e^{-\tau/\cos(\theta)}$ est le coefficient d'atténuation calculé en fonction de la profondeur optique de la couche de végétation ($\tau$) et de l'angle de vue considéré ($\theta$).

## 2.2 Correction de l'effet de la litière

### 2.2.1 Le modèle HFSS

HFSS (High Frequency Structure Simulator) est un logiciel de simulation par élément finis de la société ANSOFT. Dans ce cas précis il permet d'obtenir le coefficient de réflexion $S_{11}$, puis l'émissivité, en fonction de l'humidité pour le système (multi couches ou non) étudié. L'avantage de notre modèle HFSS repose sur la possibilité d'intégrer de nombreux paramètres comme une rugosité de surface, un gradient d'humidité ou encore la présence d'hétérogénéités dans la structure. Son inconvénient est son impossibilité d'être inversé ; c'est-à-dire que nous ne pouvons pas déduire de ce modèle de relations analytiques entre l'émissivité et les autres paramètres. Toutefois ce modèle va nous permettre de valider et d'améliorer les méthodes analytiques existantes. Les résultats ont été validés par comparaison avec d'autres méthodes.

### 2.2.2 Le modèle avec réflectivité τ-ω-R

Le principe de ce modèle est le même que celui du modèle τ-ω à la différence près que la réflectivité à l'interface litière / air n'est plus négligée. Dans ce cas on obtient une formulation en série pour chacune des trois contributions à la température de brillance du système :

Contribution de l'émission directe de la litière :
$$T_{BV1} = T_{B1}(1-\Gamma_L)\left(1 + \Gamma_S\Gamma_L\gamma_L^2 + (\Gamma_S\Gamma_L\gamma_L^2)^2 + ...\right)$$

Contribution de l'émission rétrodiffusée de la litière :
$$T_{BV2} = T_{B1}(1-\Gamma_L)\Gamma_S\gamma_L\left(1 + \Gamma_S\Gamma_L\gamma_L^2 + (\Gamma_S\Gamma_L\gamma_L^2)^2 + ...\right)$$

Contribution de l'émission directe du sol :
$$T_{BS} = T_S(1-\Gamma_L)(1-\Gamma_S)\gamma_L\left(1 + \Gamma_S\Gamma_L\gamma_L^2 + (\Gamma_S\Gamma_L\gamma_L^2)^2 + ...\right)$$

Nous obtenons l'écriture suivante pour la température de brillance :

$$T_B = \frac{1-\Gamma_L}{1-\Gamma_L\Gamma_S\gamma}\left[(1+\gamma\Gamma_S)(1-\gamma)(1-\omega)T_L + (1-\Gamma_S)\gamma T_S\right]$$

où   l'indice L fait référence à la litière et l'indice S fait référence au sol.

$\Gamma_S$ et $\Gamma_L$ sont respectivement la réflectivité à l'interface sol / litière et à l'interface litière / air ;

$T_S$ et $T_L$ sont respectivement les températures du sol et de la litière ;

$\gamma = e^{-\tau/\cos(\theta)}$ est le coefficient d'atténuation.

## 3   Résultats

Le site expérimental de l'INRA du Bray (33) est équipé d'une tour de 40 m de haut au sommet de laquelle un radiomètre permet d'effectuer des mesures sur la forêt environnante. Ces dernières ont été effectuées pour un angle θ variant entre 25° et 60°, en polarisation H.

Les mesures montrent que l'émissivité du sol (qui à ce stade de l'étude est constitué de terre et de litière) dépend non seulement de l'humidité de la terre mais également de celle de la litière. Il nous reste à déterminer quelle est la configuration de cette couche de litière puis à la modéliser de manière simple pour pouvoir corriger ses effets.

Les mesures ont également permis de mettre en évidence le lien qu'il existe entre l'humidité volumétrique du sol (SM) et l'humidité gravimétrique de la litière (LM). Il est alors possible de déduire une relation affine entre les deux humidités :

$$LM (grav. \%) = 2.7201\ SM (vol. \%) - 8.6223$$

Cette courbe a été déduite des mesures réalisées sur une période de 6 mois et l'approximation est juste avec un facteur de corrélation de 0.8593.

Avant de pouvoir comparer les mesures aux résultats de simulations il est nécessaire d'effectuer une correction de l'émissivité en fonction de l'angle d'incidence du radiomètre lors des mesures. En effet, sous HFSS les calculs sont effectués au nadir (θ=0) tandis que les mesures considérées ont été effectuées à θ=45°.

Les mesures ayant été effectuées pour différents angles d'incidence (de 25° à 60°), nous considérons une température de brillance moyenne pour chacun d'entre eux. Nous avons ensuite approximé ces mesures à une courbe de la forme :

$$T_B = A\left(1 - e^{\frac{-(90-\theta)}{d}}\right)$$

Où A et d sont des coefficients à fixer en fonction des mesures effectuées sur le site du Bray.

Des caractérisations ont été effectuées [4] afin d'évaluer la permittivité de la terre et de la litière du site expérimental du Bray. Un domaine de permittivité a ainsi été défini pour tenir compte des erreurs de mesure et de leur répétabilité (liée à la non homogénéité du matériau).

Dans un deuxième temps nous avons calculé l'émissivité équivalente du système terre + litière à l'aide du modèle HFSS pour différentes épaisseurs de litière. La couche de litière a dans la réalité une épaisseur variable selon la topologie du sol. Pour tenir compte de ce paramètre nous allons prendre comme courbe d'émissivité équivalente une courbe moyennant les émissivités de la structure terre + litière pour des épaisseurs de litière variant de 2 à 6 cm.

Nous avons calculé ensuite la correction par le modèle τ−ω (suppression de l'effet de la végétation) sur les mesures radiométriques expérimentales (courbe noire sur la Figure 1). Les résultats des deux méthodes sont présentés sur la Figure 1. L'épaisseur de litière retenue est en moyenne de 4cm. Un domaine d'émissivité (en gris sur la figure) est défini pour tenir compte des erreurs sur les mesures de permittivité introduite dans HFSS [4]. Nous pouvons constater le bon accord entre les simulations et les mesures corrigées. L'étape suivante consiste à développer un modèle analytique simple qui puisse être intégré à l'algorithme de calcul SMOS et qui permettra la correction de l'effet de la litière.

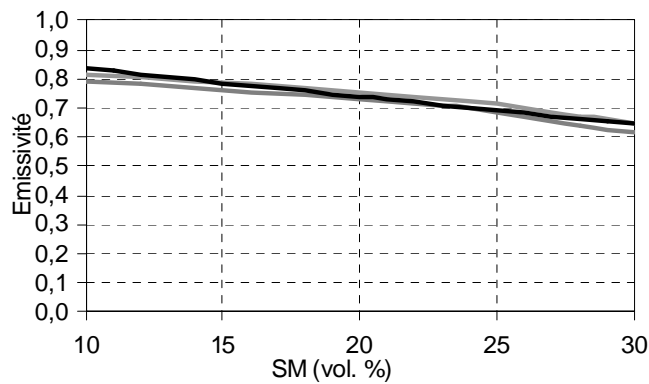

**Figure 1 : émissivité du système terre + litière en considérant une épaisseur moyenne de 4 cm de litière**

Le modèle τ-ω néglige la réflectivité de la couche dont nous voulons corriger l'effet sur l'émissivité. Cette hypothèse est fondée pour la couche de végétation dont la réflectivité est faible. Appliquer le modèle τ-ω à la correction de l'effet de la litière reviendrait à considérer que la réflectivité de la couche de litière est négligeable. Les mesures effectuées sur la litière contredisent cette hypothèse.

L'étape suivante consiste à déterminer dans quelle mesure le modèle analytique permet de calculer l'émissivité du sol sachant que $\Gamma_S$, $\Gamma_L$, $T_S$, $T_L$ et l'épaisseur de la litière sont connus tandis que τ est à calculer. Pour les mêmes entrées nous avons comparé les émissivités de la terre seule issues du modèle HFSS et τ−ω-R. Un domaine d'émissivité est défini pour tenir compte des erreurs sur les mesures de permittivité introduite dans

HFSS [4]. La Figure 2 présente les résultats obtenus en considérant une couche de litière d'épaisseur moyenne de 4 cm.

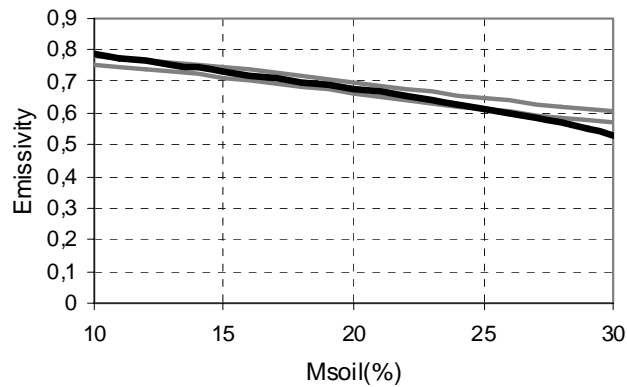

**Figure 2 : Emissivité du sol par la méthode τ-ω en noir et emissivité du sol via HFSS en gris**

Là encore la bonne concordance des résultats nous encourage quand au choix du modèle τ−ω-R pour corriger l'effet de la litière.

### 4   Conclusions et perspectives

Le logiciel de simulation numérique HFSS nous a permis de mettre au point un modèle de calcul d'émissivité de sol. Les simulations ont mis en évidence l'influence à la fois non négligeable et non triviale de la litière sur l'émissivité équivalente de la surface. Après avoir utilisé un modèle d'inversion de type τ−ω pour corriger l'effet de la végétation nous avons vérifié la pertinence d'un modèle τ−ω-R pour corriger l'effet de la litière. La bonne concordance des résultats issus de différents modèles a conforté notre choix.

La prochaine étape de nos travaux va consister à étudier l'évolution des propriétés électromagnétiques des milieux concernés (terre et litière) au cours d'une année (influence des saisons…) et de calculer l'effet de ces variations sur la mesure de l'émissivité équivalente de la structure géologique.

Par la suite nous étudierons l'influence de la rugosité de surface du sol sur l'émissivité. Dans cette optique, l'influence de la litière couplée à la rugosité du sol sera abordée. Des termes correctifs pourraient donc être intégrés à notre modèle d'inversion τ−ω−R.

Nous souhaitons aussi étudier l'émissivité équivalente de surfaces plus importantes qui représenteraient davantage la réalité. Outre la rugosité ou l'épaisseur de litière, nous introduirons des gradients d'humidité afin d'étudier l'émissivité globale équivalente.